\documentclass{article}

\usepackage{arxiv}

\usepackage[utf8]{inputenc} 
\usepackage[T1]{fontenc}    
\usepackage{hyperref}       
\usepackage{url}            
\usepackage{booktabs}       
\usepackage{amsfonts}       
\usepackage{nicefrac}       
\usepackage{microtype}      
\usepackage{graphicx}
\usepackage[numbers]{natbib}
\usepackage{doi}

\usepackage{graphicx}
\usepackage{arxiv}
\usepackage{acro}

\usepackage[utf8]{inputenc} 
\usepackage[T1]{fontenc}    
\usepackage{setspace}
\usepackage{hyperref}       
\usepackage{url}            
\usepackage{booktabs}       
\usepackage{amsfonts}       
\usepackage{nicefrac}       
\usepackage{microtype}      
\usepackage{lipsum}         
\usepackage{natbib}
\usepackage{doi}

\usepackage{subcaption}

\usepackage{algorithm} 
\usepackage{algorithmic}  
\usepackage[algo2e]{algorithm2e}

\usepackage[acronym]{glossaries}
\usepackage{booktabs}

\usepackage{textcomp}
\usepackage{xspace}

\usepackage{amsmath} 
\usepackage{setspace}

\usepackage{placeins} 

\usepackage{enumitem}
\usepackage{balance}
\usepackage{lineno}

\input{commands}

\DeclareAcronym{cvt}{
  short=CVT,
  long=centroidal {V}oronoi tessellation,
}
\DeclareAcronym{Cryo-em}{
  short=cryo-EM,
  long=cryogenic electron microscopy ,
} 
\DeclareAcronym{Cryo-et}{
  short=cryo-ET,
  long=cryogenic electron tomography,
} 
\DeclareAcronym{sssr}{
  short=SSSR,
  long=simple and scalable surface reconstruction,
}
\DeclareAcronym{mat}{
  short=MAT,
  long=medial axis transform,
}
\DeclareAcronym{SDF}{
  short=SDF,
  long=Signed Distance Field,
}
\DeclareAcronym{FIB}{
  short=FIB,
  long=focused ion beam,
}
\DeclareAcronym{SEM}{
  short=SEM,
  long=scanning electron microscopy
}
\DeclareAcronym{FIB-SEM}{
  short=\acs{FIB}-\acs{SEM},
  long=\acl{FIB}-\acl{SEM}
}
\DeclareAcronym{PSR}{
  short=PSR,
  long=Poisson Surface Reconstruction,
}
\DeclareAcronym{MD}{
  short=MD,
  long=molecular dynamics,
}
   
\graphicspath{{figs/}{figures/}{pictures/}{images/}{./}} 

\title{MidSurfer: A Parameter-Free Approach for  Mid-Surface Extraction from Segmented Volumetric Data}

\author{%
    \href{https://orcid.org/0000-0002-3181-3018}{\includegraphics[scale=0.06]{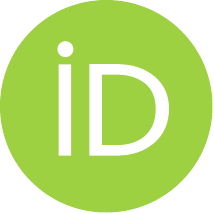}\hspace{1mm}Eva Bone\v{s}$^{1,2,*}$},
    \href{https://orcid.org/0000-0001-5864-1888}{\includegraphics[scale=0.06]{orcid.pdf}\hspace{1mm}Dawar Khan$^{1,*}$},
    \href{https://orcid.org/0000-0002-9015-2897}{\includegraphics[scale=0.06]{orcid.pdf}\hspace{1mm}Ciril Bohak$^{2}$}, 
    \href{https://orcid.org/0000-0002-9015-2897}{\includegraphics[scale=0.06]{orcid.pdf}\hspace{1mm}Benjamin A. Barad$^{3}$},\And  
    \href{https://orcid.org/0000-0002-9015-2897}{\includegraphics[scale=0.06]{orcid.pdf}\hspace{1mm}Danielle A. Grotjahn$^{4}$},
    \href{https://orcid.org/0000-0001-5908-7882}{\includegraphics[scale=0.06]{orcid.pdf}\hspace{1mm}Ivan Viola$^{1}$} and
    \href{https://orcid.org/0000-0001-7907-7382}{\includegraphics[scale=0.06]{orcid.pdf}\hspace{1mm}Thomas Theu\ss l$^{1}$}.
    \\
       $^{1}$King Abdullah University of Science and Technology (KAUST), Saudi Arabia.\\
        E-mail: \{eva.bones\,$|$\,dawar.khan\,$|$\,ivan.viola\,$|$\,thomas.theussl\}@kaust.edu.sa. \\
        $^{2}$University of Ljubljana, Slovenia.
        E-Mail: ciril.bohak@fri.uni-lj.si. \\
        $^{3}$Oregon Health \& Science University, OR, United States. 
        E-mail: barad@ohsu.edu.\\
        $^{4}$The Scripps Research Institute, CA, United States.
        E-mail: grotjahn@scripps.edu.\\
        $^{*}$E. Bone\v{s} and D. Khan are co-first authors.
}
 
\date{}

\renewcommand{\shorttitle}{MidSurfer: Mid-Surface Extraction from Segmented Volumetric Data}

\begin{document}
\maketitle

\begin{figure}[!h]
    \centering
\vskip -2.0cm
  \includegraphics[width=\linewidth]{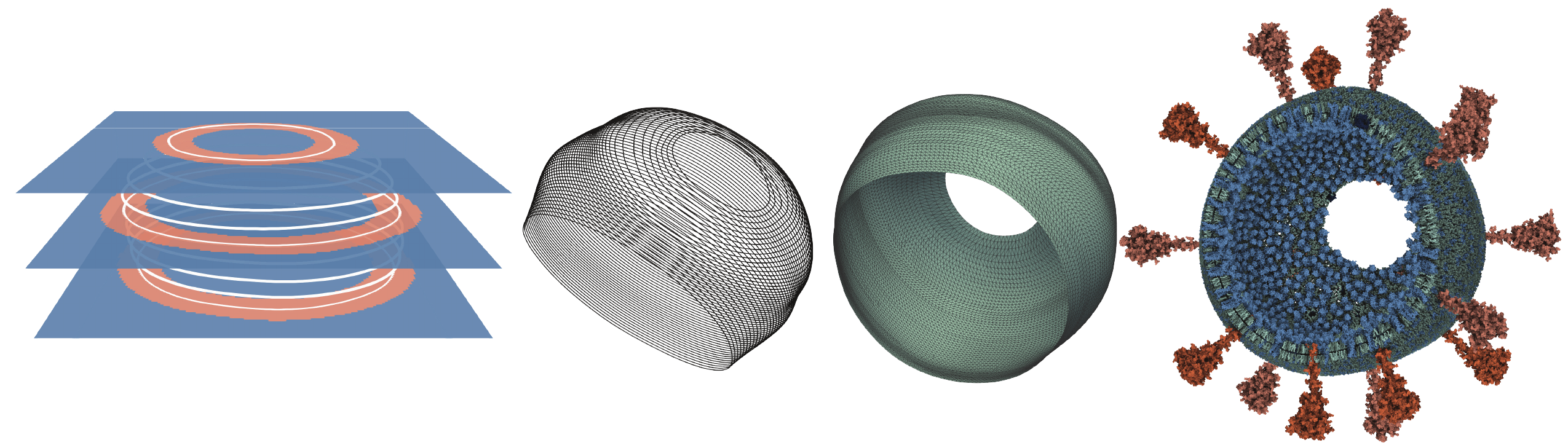}
  \caption{%
     An overview of the overall method: From left to right: Volumetric segmented data represented in slice-wise view, the extracted mid-polylines from the data, the triangular surface generated from mid-polylines, and a use case of the mid-surface with the modeling results visualized over the extracted mid-surface mesh.
  }
  \label{fig:teaser}
\end{figure}

\begin{abstract}
In the field of volumetric data processing and analysis, extracting mid-surfaces from thinly bounded compartments is crucial for tasks such as surface area estimation and accurate modeling of biological structures, yet it has lacked a standardized approach. 
To bridge this gap, we introduce MidSurfer--a novel parameter-free method for extracting mid-surfaces from segmented volumetric data.
Our method produces smooth, uniformly triangulated meshes that accurately capture the structural features of interest. 
The process begins with the Ridge Field Transformation step that transforms the segmented input data, followed by the Mid-Polyline Extraction Algorithm that works on individual volume slices. 
Based on the connectivity of components, this step can result in either single or multiple polyline segments that represent the structural features. 
These segments form a coherent series across the volume, creating a backbone of regularly distributed points on each slice that represents the mid-surface. 
Subsequently, we employ a Polyline Zipper Algorithm for triangulation that connects these polyline segments across neighboring slices, yielding a detailed triangulated mid-surface mesh. 
Our findings demonstrate that this method surpasses previous techniques in versatility, simplicity of use, and accuracy. 
Our approach is now publicly available as a plugin for ParaView, a widely-used multi-platform tool for data analysis and visualization, and can be found at (\href{https://github.com/kaust-vislab/MidSurfer}{\small{\url{https://github.com/kaust-vislab/MidSurfer}}}).
 
\end{abstract}


\section{Introduction}
\maketitle
In the exploration of biological structures, the visualization and analysis of surfaces play a crucial role in understanding complex biological phenomena. In particular, the study of thinly bounded compartments, such as cells and their organelles bounded with membranes, requires precise surface representation for accurate analysis since membranes serve as interfaces for various cellular components such as cell membranes, protein-protein interfaces, organelle junctions, and cell-extracellular matrix contacts. Membranes govern molecular exchange, cellular communication, and intracellular transport, thereby shaping cellular function and organization. These structures, often explored through volumetric microscopy data (obtained using \ac{Cryo-et}, \ac{FIB-SEM}, or some other acquisition method), contain rich details critical for various biological research and applications. However, the extraction and analysis of surfaces from such data pose significant challenges due to the intricate nature of biological structures and the limitations of existing approaches.

The mid-surface concept~\cite{barad2023quantifying} serves as a crucial tool in this context, effectively bridging the divide between raw volumetric data and a meaningful geometric representation of biological structures. Broadly defined, a mid-surface~\cite{rezayat1996midsurface, kulkarni2017leveraging} is a surface model that captures the structure's geometric essence by depicting its median or central layer, which is equidistant from the object's inner and outer surfaces. This model is invaluable for analyzing thinly bounded compartments, offering a well-balanced and precise surface for subsequent analysis. Notably, the mid-surface, situated between the inner and outer leaflets of the lipid bilayer, is of particular importance for understanding membrane structure and various cellular functions, such as membrane fluidity, protein localization, and cell signaling. 

Distinguishing between mid-surfaces, medial surfaces~\cite{price1997hexahedral}, and isosurfaces~\cite{keppel1975approximating}, illustrated in~\cref{fig:surfaces}, is crucial, as these concepts, while often related, serve distinct purposes in volumetric data analysis. Mid-surfaces are designed to represent the central geometry of thinly bounded spaces, capturing the essence of an object's median layer. Medial surfaces, in contrast, focus on the geometric and topological core of a shape, essentially forming its skeleton. Isosurfaces, meanwhile, are defined by extracting surfaces that represent a constant value within a scalar field, a technique commonly employed in volume rendering. Additionally, the concept of outer surfaces~\cite{feng2012geometric,lindow2014ligand, salfer2020reliable}, including molecular surfaces~\cite{liu2015parameterization} and membrane surfaces~\cite{sadeghi2018particle}, plays a significant role in various analysis scenarios. Among these, mid-surfaces are particularly advantageous for the detailed analysis of biological structures due to their unique ability to accurately represent the median geometry. However, it is noteworthy that, to the best of our knowledge, the scientific literature primarily mentions a single approach~\cite{barad2023quantifying} for extracting mid-surfaces, which necessitates adjusting multiple parameters to achieve the desired outcome. Other methods for extracting similar surfaces, as discussed in \autoref{sec:related-work}, are either not fit for purpose or fail to ensure the topological integrity of a manifold with a boundary.

In biological research, precise surface representation is critical for analyzing and modeling cellular compartments, necessitating new surface extraction methods. Traditional approaches often fall short in capturing the intricate details required for thorough biological analyses, particularly from volumetric microscopy data. The mid-surface concept, representing the median layer of an object and providing an accurate model for thinly bounded structures, emerges as a solution. However, the literature indicates a significant gap: a standardized, parameter-free method for mid-surface extraction is notably absent, with existing methods requiring manual parameter adjustments and often failing to maintain topological integrity. This gap underscores the need for an innovative approach that combines accuracy with usability in extracting mid-surfaces from biological data. Our work addresses this necessity, proposing a novel, efficient method that enhances the geometric representation of biological structures, thereby facilitating a deeper understanding of their complex functions.


 The key contributions of the presented method include:

\begin{itemize}[nolistsep]
\item Introducing a novel, parameter-free, and robust method for extracting the mid-surface from segmented volumetric data.
\item Representing the mid-surface in the form of a high-quality triangular surface mesh, while considering the standard mesh quality metrics.
\item Evaluating the method across multiple datasets, demonstrating significant improvements over existing alternatives. Additionally, showcasing its efficacy in two structural biology use cases, including quantitative analysis with surface morphometrics and mid-surface modeling.
\item Making our algorithm publicly available as a plugin in ParaView\footnote{\url{https://www.paraview.org}}, an open-source visualization tool, thereby providing accessible functionality to a wide range of users, from novices to experts.
\end{itemize}

\section{Related Work}
\label{sec:related-work} 
In 3D modeling, surfaces serve as foundational structures across various applications, facilitating the attainment of desired final shapes. These surfaces may be manually crafted by users or algorithmically extracted from reference data, encompassing a wide array of geometric representations, including meshes, point clouds, and volumetric data. Historically, numerous approaches for surface extraction have been developed, primarily influenced by the nature of the input data. However, it's imperative to approach these methods with flexibility, as input data can often be converted among different formats with high precision. Consequently, we review related works capable of processing various types of input data.

Point cloud data, one of the prevalent forms of representation, poses unique challenges and opportunities for surface reconstruction. The goal is either to fit a surface that encompasses all points or to devise a surface model that optimally represents the point cloud. Pioneering contributions in this area were made by Hoppe~\etal~\cite{Hoppe1992, Hoppe1993}, who introduced innovative techniques for the surface approximation of unorganized point sets. Boissonnat and Cazals~\cite{Boissonnat2000} offered a method for smooth surface reconstruction from arbitrary point distributions using Natural Neighbour interpolation of the distance function. Approaches to simplify surfaces derived from unstructured point clouds were advanced by Pauly~\etal~\cite{Pauly2002, Pauly2003}. Subsequent research introduced methods for fitting smooth surfaces to arbitrary shapes using moving least squares combined with local mappings or other features~\cite{Alexa2001, Alexa2003, Levin2004, Amenta2004}, with further developments aimed at preserving sharp features~\cite{Fleishman2005, Guennebaud2007, Oeztireli2009}. Ohtake~\etal~\cite{Ohtake2005} demonstrated surface fitting excellence using multi-level partitioning of unity implicits. The advent of deep neural networks has also made a significant impact on surface reconstruction, achieving remarkable outcomes. Williams~\etal~\cite{Williams2019} explored the use of deep neural networks as a geometric prior in this context, while Boulch and Marlet~\cite{Boulch2022} tackled the scalability challenges inherent in deep learning-based surface reconstruction techniques. Comprehensive reviews of these advancements are documented in survey papers by Berger~\etal~\cite{Berger2017} and Huang~\etal~\cite{Huang2022}.

It is worth noting that many of the above-mentioned methods are adaptable to volumetric data through a straightforward conversion of volumes to point clouds, as illustrated by some approaches, including~\cite{Ohtake2005}, which demonstrate promising results. However, these techniques are primarily designed for unstructured data, contrasting with the highly regular data characteristic of our use cases. By leveraging the regularity inherent in our datasets, our method achieves greater efficiency and alignment with our specific requirements.

\begin{figure}
    \centering
   \includegraphics[width=0.8\linewidth]{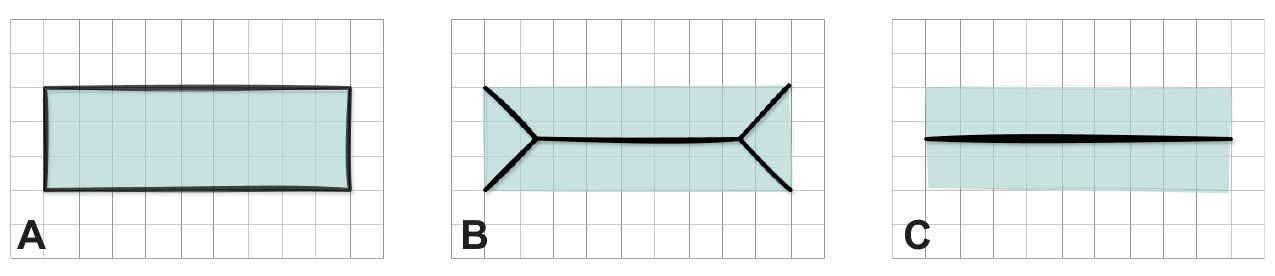}
   \caption{Comparative illustration of (A) isosurfaces, (B) medial surfaces, and (C) mid-surfaces to emphasize the unique advantages of mid-surfaces in capturing the median geometry. The segmentation is shown in light green color and the respective surfaces with a thick black line.}
  \label{fig:surfaces}
\end{figure}

While point cloud data is frequently associated with surface reconstruction, the challenges inherent in working with volumetric data stem from its continuous representation of structures. Unlike point clouds, where the goal is often to reconstruct discrete surfaces or a collection of interconnected surfaces, volumetric data requires an initial determination of what constitutes a surface within the volume. Typically, this classification divides surfaces into three primary categories as illustrated in~\cref{fig:surfaces}: (1) \emph{isosurfaces}~\cite{keppel1975approximating}, which delineate boundaries within volumes of uniform value; (2) \emph{medial surfaces}~\cite{blum1967t} (also known as \emph{medial axes}), representing the structural skeleton; and (3) \emph{mid-surfaces}, situated equidistantly between two boundary surfaces, such as isosurfaces or the edges of a segmented region. Our research primarily addresses the latter category, though we acknowledge key contributions to the former two due to their relevance and impact on our work.

The Marching Cubes algorithm~\cite{lorensen1987marching}, celebrated for its simplicity and efficiency, remains the quintessential method for isosurface extraction, leveraging linear interpolation along voxel edges. Subsequent attempts to resolve its ambiguities or improve upon its drawbacks, such as those by~\cite{Doi1991, Treece1999}, have not supplanted it, often due to increased complexity and reduced performance. Efforts to generate closed and oriented surfaces through valid triangulation include Guzetic~\etal's~\cite{Gueziec1995} tetrahedral decomposition-based method. Livnat~\etal~\cite{Livnat1996} proposed a near-optimal isosurface extraction technique for both structured and unstructured grids using span space representation. Wood~\etal~\cite{Wood2000} introduced an approach tailored for distance volumes, beginning with a coarse mesh extraction that is refined via a multiscale force-based solver to achieve a semi-regular mesh with adaptive geometric sampling. To mitigate volume sampling aliasing, Kobbelt~\etal~\cite{Kobbelt2001} developed a feature-sensitive surface extraction method that preserves the simplicity of Marching Cubes. Ju~\etal~\cite{Ju2002} presented an efficient, real-time capable octree-based approach utilizing Hermite contouring, which does not necessitate the explicit identification and processing of specific features. Addressing the precision required for medical surface extractions, Kim~\etal~\cite{Kim2005} demonstrated an approach using a Laplacian map, particularly for extracting cortical surfaces in the human brain. The advent of differentiable methods for isosurface extraction, aimed at facilitating rapid optimization with physically informed neural networks, led to novel approaches like that of Remelli~\etal~\cite{Remelli2020}, which employs Deep Signed Distance Functions for explicit surface mesh representation, demonstrated in single-view reconstructions and physically-driven shape optimization.


Although the methods described above do not directly address our specific challenges, theoretically, our problem could be reformulated as an isosurface extraction task in some specific scenarios. However, this is not applicable to many of our cases due to the complexity of the structures we aim to address.

Medial surfaces, often depicted through medial meshes~\cite{yong2019multi}, encapsulate an object's core geometric and topological features~\cite{yan2018voxel}. Yet, in the context of our work, medial surfaces prove inadequate due to the complexity of the structures we examine. The skeletons of these complex structures are typically branched and fail to provide a practical foundation for mesoscale modeling.

This limitation underscores the necessity for a mid-surface representation~\cite{rezayat1996midsurface}, a concept widely applied in fields such as engineering and finite element analysis~\cite{woo2014abstraction}, as well as structural biology, as highlighted by Barad~\etal~\cite{barad2023quantifying}. In contrast to isosurfaces, which excel in depicting uniform structures by showcasing constant values or medial surfaces that reduce shapes to their skeletal essence, mid-surfaces uniquely capture the median layer situated between two boundaries. This distinct trait renders mid-surfaces particularly adept for the detailed geometric exploration of thinly bounded structures.

Barad~\etal's approach employs a mid-surface morphometrics pipeline alongside meshing techniques to model and quantify various metrics, including curvature and distances within and between surfaces, in addition to orientation differences. Their methodology integrates several existing techniques, each with specific input parameters to yield optimal outcomes. They adopted a semi-automated algorithm~\cite{martinez_sanchez2014robust} for segmenting cellular membranes from \ac{Cryo-et} data, followed by the use of screened \ac{PSR}~\cite{kazhdan2013screened} for mesh generation. While \ac{PSR} effectively transforms membrane voxel segmentations or point clouds into implicit surface meshes, there's a notable risk of geometrical fidelity loss and the obliteration of sharp features and voids unless input parameters are meticulously chosen. Despite these challenges, the quantitative findings of Barad~\etal provide a robust solution to a complex problem. Nonetheless, the quest for optimal parameter settings remains daunting. Their investigations further highlight the demand for more automated, parameter-free, and structurally reliable methods for mid-surface extraction—our research's primary aim.

In one phase of our method, we employ a technique to seamlessly merge parallel contours into a unified mesh representation, a process known as \emph{zippering}, the Zipper algorithm, or contour stitching. Various strategies for addressing this challenge in a broader context have been documented in prior works~\cite{keppel1975approximating,fuchs1977optimal,christiansen1978conversion}. Our approach is particularly influenced by the method developed by Turk and Levoy~\cite{turk1994zippered}, who presented the method for the creation of zippered polygon meshes. Their algorithm focuses on amalgamating multiple mesh fragments obtained from different viewpoints into a cohesive mesh that represents a singular object. Unlike their mesh-to-mesh algorithm, our implementation is tailored to converting stacked polylines into mesh structures. Furthermore, our method innovatively incorporates new triangles to facilitate connections between these structures—a modification not explicitly required in Turk and Levoy's original formulation. It is also worth noting the existence of Zipper~\cite{GURUNG2013262}, which introduces a mesh data structure aimed at efficient mesh representation. However, this does not directly pertain to the mesh generation.

Another phase of our method extracts mid-polylines from individual slices, a problem that is closely related to ridge extraction as presented by Furst~\etal~\cite{Furst1996} and by Furst and Pizer with Marching Ridges~\cite{Furst2001}, where ridges are defined as level sets of first derivatives of the input function. This has been subsequently used, \eg by Sadlo and Peikert~\cite{sadlo2009visualizing} for visualizing Lagrangian coherent structures extracted as ridges from FTLE fields, and Kindlmann~\etal~\cite{kindlmann2006anisotropy} for identifying white matter structures in Diffusion Tensor MRI.



\section{Method}
\label{sec:method}
\begin{figure}
    \centering
    \includegraphics[width=0.8\linewidth]{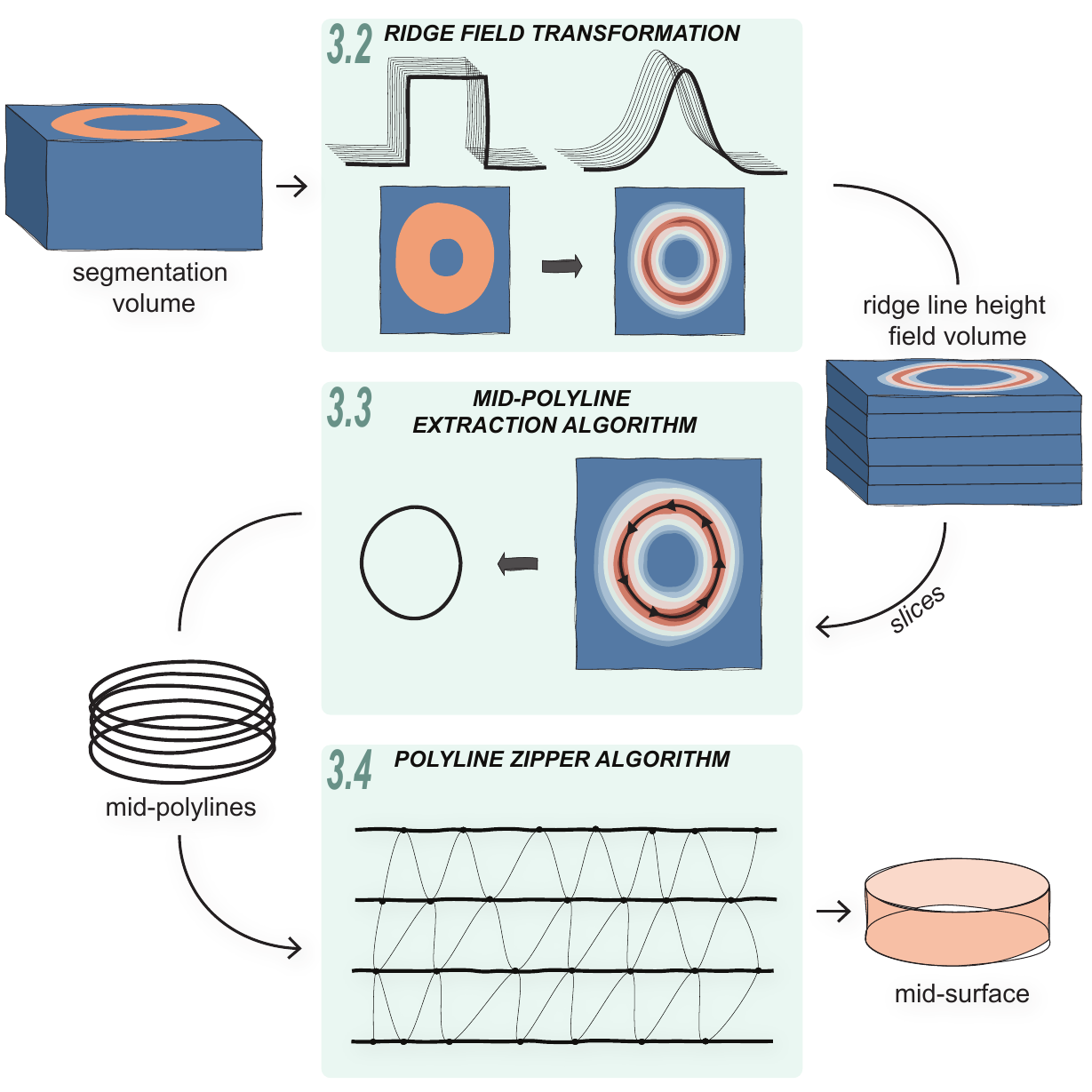}
    \caption{
    Schematic overview of the Mid-surface Extraction Algorithm, showing steps from Ridge Field Transformation (\cref{sec:ridge_field_transformation}) that transforms the binary segmentation data; through the Mid-Polyline Extraction Algorithm (\cref{sec:centerLines} that extracts the mid-polylines from slices; to the final mesh generation using the Polyline Zipper Algorithm (\cref{sec:meshing}).}
    \label{fig:method}
\end{figure}

Our method processes binary segmentations of volumetric data to derive mid-surfaces from individually segmented structures. Two critical definitions underpin our approach:

\paragraph{Mid-polyline} is defined as a series of straight line segments that lie precisely midway between the inner and outer boundaries of a segmentation, confined to a single slice. This principle underlies the Mid-polyline Extraction Algorithm detailed in~\cref{sec:centerLines}, serving as its foundational concept.

\paragraph{Mid-surface} is conceptualized as a two-dimensional manifold centrally embedded within a three-dimensional structure, equidistant from the structure's inner and outer surfaces. The transformation from mid-polylines to cohesive mid-surfaces is achieved through our Polyline Zipper Algorithm, streamlining the construction of the desired surfaces.

\subsection{Method Overview}
\cref{fig:method} illustrates the overall pipeline, encompassing three key modules detailed in the subsequent subsections. The overarching goal is to accurately extract a mid-surface from volumetric data and represent it as a triangular surface mesh. This mesh is constructed by triangulating a set of horizontally aligned mid-polylines (\cref{sec:meshing}). These mid-polylines are derived from segmented volumetric data (\cref{sec:centerLines}). The volumetric data is analyzed slice-by-slice, with each slice yielding mid-polylines at its specific level. When the segmentation within a slice is continuous, it results in a single, seamless mid-polyline. Conversely, discontinuities in the segmentation lead to fragmented mid-polyline segments at that level.
To derive a mid-polyline from a given slice, we first convert the binary, cliff-like segmentation into a gradient, hill-range-like height field with a ridge in the center that we term a \emph{ridge line height field} (\cref{sec:ridge_field_transformation}).
This conversion process emphasizes symmetry to guarantee the central ridge line accurately delineates the mid-polyline. The symmetry of the transformation ensures the extracted ridge line precisely embodies the mid-polyline's path. We initiate the extraction process by identifying the highest point on the ridge line height field, which serves as our starting position. From there, the method proceeds in the ridge's direction, which aligns with the direction of minimal curvature. This determination is based on the analysis of the Hessian matrix, which compiles the second-order derivatives for each pixel within the slice. By calculating the matrix's eigenvectors and eigenvalues, we ascertain the primary curvatures, guiding our extraction process along the minimal curvature direction.

\subsection{Ridge Field Transformation} 
\label{sec:ridge_field_transformation}
To successfully extract the mid-polyline in the following stages, it is imperative to generate a smooth height field featuring a central ridge, known as a ridge line height field. This transformation must be symmetric, ensuring the ridge is precisely positioned midway between the segmentation boundaries. Additionally, it should seamlessly cover the entire breadth of the segmented area, facilitating an accurate and comprehensive extraction of the mid-polyline.

Na\"ively, a method that meets these requirements could be a convolution operator that performs a weighted average of surrounding pixels. This is often achieved through Gaussian smoothing, defined as:
\begin{equation}
G(x, y, z) = \frac{1}{{(2\pi\sigma^2)^{3/2}}} e^{-\frac{x^2 + y^2 + z^2}{2\sigma^2}} ,
\end{equation}
where \( G(x, y,z) \) represents the Gaussian kernel, \( \sigma \) denotes the standard deviation (or kernel radius) of the Gaussian distribution, and \( (x, y, z) \) indicate kernel position. 
Choosing the proper parameters, particularly the standard deviation, is critical for achieving accurate results. The optimal parameters depend on the width of the segmentation. For wider segmentations, larger smoothing values are required to ensure a smooth transition across the entire area, preventing leaving any constant patches. Conversely, narrower segmentations demand smaller values to avoid overly diminishing the features within the resulting height field.

The challenge of manually setting the accurate parameters---and the need to apply a single parameter set across the entire volume, which might contain segmentations of varying widths---prompted us to explore a different approach. 
Drawing on the understanding that the transformation should depend on the segmentation's width and the previous requirements for symmetry and a smooth ridge line height field, we utilize a \ac{SDF}.
\Ac{SDF} assigns a distance value to each pixel/voxel relative to the nearest boundary. Mathematically, the \ac{SDF} can be defined as follows:
\begin{equation}\label{eq:SDF}
\text{SDF}(x, y, z) = \min(d_{\text{in}}(x, y, z), d_{\text{out}}(x, y, z)),
\end{equation}
where, $d_{\text{in}}(x, y, z)$ represents the distance to the inner boundary, while $d_{\text{out}}(x, y, z)$ represents the distance to the outer boundary. 
This formulation ensures that points lying exactly at the midpoint between the two boundaries have the highest values. Consequently, it yields a ridge line height field where the highest points are located in the middle between the two boundaries, irrespective of their distance (\ie the thickness of the segmentation).

The \ac{SDF} accurately identifies the center of the segmentation. However, the resolution of the original segmentation influences the resulting \ac{SDF} and can introduce a staircasing effect. 
To counteract this issue, we apply smoothing to the \ac{SDF} using the Gaussian kernel. However, to keep the approach parameter-free and adapt to the specific dataset in use, the smoothing parameters for this step are also derived from the \ac{SDF}, where the standard deviation is half of the \ac{SDF} maximum value $\sigma = \frac{SDF_{max}}{2}$ and the size of the kernel is calculated accordingly to reasonably fit the Gaussian curve $radius = 2\sigma + 1$. Gaussian kernel $G_s(x, y, z)$ in our case is defined as: 
\begin{equation}
G_s(x, y, z)= \frac{1}{{(2\pi\left(\frac{SDF_{\text{max}}}{2}\right)^2)^{3/2}}} e^{-\frac{x^2 + y^2 + z^2}{2\left(\frac{SDF_{\text{max}}}{2}\right)^2}}.
\label{eq:gaussian_smoothing2}
\end{equation}
The final smoothed \ac{SDF} is obtained by convolving, denoted by $*$, the original \ac{SDF} (\cref{eq:SDF}) with the Gaussian smoothing (\cref{eq:gaussian_smoothing2}):
\begin{equation}
\text{SDF}_{\text{smoothed}}(x, y, z) = \text{SDF}(x, y, z) * G_s(x, y, z).
\label{eq:SDF_smoothing2}
\end{equation}

The intended outcome of this process is a uniform ridge line height field across slices. 
To achieve this, we apply the procedure to the entire volume, ensuring consistent transitions between slices and eliminating the staircasing effect typically observed in slice-based 2D processing. 
This means that both the calculation of the \ac{SDF} and its subsequent smoothing are performed in 3D.

For slicing, we consider the axial orientation of the microscopy data, which comes in a stacked format with an up-and-down direction based on sample preparation. This inherent directionality guides our slicing direction to ensure it aligns with the natural structure and preparation of the sample. The axial slices are known to have the highest resolution due to the limited range of the tilting process during the microscopy data acquisition.

\begin{figure}[b]
    \centering 
    \includegraphics[width=0.8\linewidth]{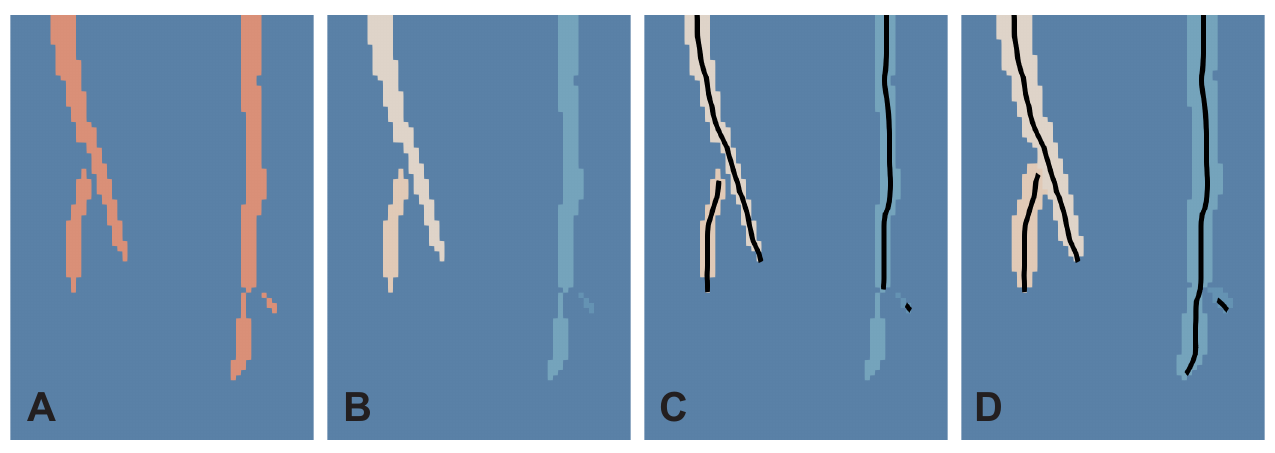}
    \caption{
    Visualization of the connected components identification and mid-polyline tracing on a binary mask slice. (A) The original binary mask. (B) The connected components within the binary mask, each assigned a unique ID. (C) Tracing the mid-polyline within these identified connected components. (D) Tracing the mid-polyline on the dilated connected components, avoiding ending the line prematurely around narrow boundaries.}
    \label{fig:dilation}
\end{figure}

\subsection{Mid-Polyline Extraction Algorithm}
\label{sec:centerLines} 

\begin{figure*}[tb]
    \centering 
    \includegraphics[width=\textwidth]{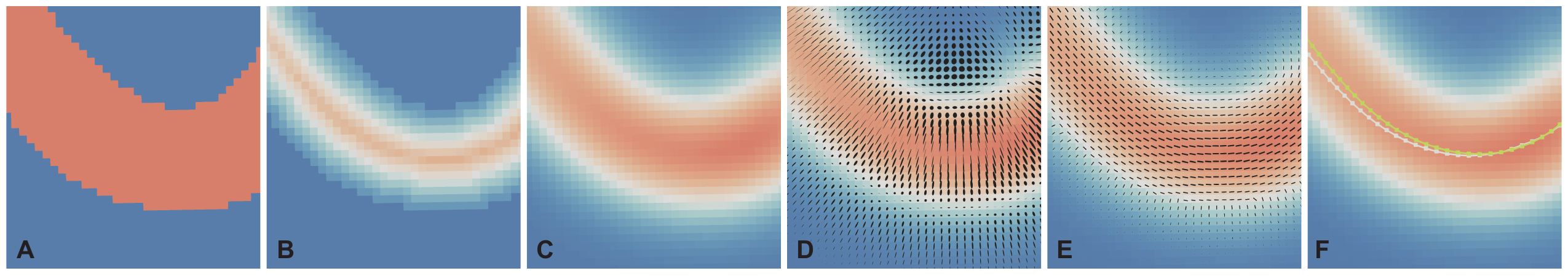}
    \caption{Visualization of Ridge Field Transformation and Mid-Polyline Extraction Algorithm on a slice:
    (A) Initial binary segmentation, showcasing the structure of interest.
    (B) Signed distance field derived from the initial segmentation.
    (C) Ridge line height field (a smoothed version of the signed distance field to ensure a consistent non-zero derivative across all pixels).
    (D) Curvature tensor field visualization, derived from the ridge line height field, using superquadric tensor glyphs~\protect\cite{kindlmann2004superquadric}.
    (E) Visualization of normalized eigenvectors associated with the smallest eigenvalue, represented as a line field. 
    (F) Golden section search optimization: a white polyline depicts a streamline within an eigenvector field drifting from the ridge, while a green line shows the golden section search perpendicular to the eigenvector, ensuring the streamline's alignment with the ridge.}
    \label{fig:mask_to_line_field_2}
\end{figure*}

Having a stack of slices (each in 2D) with a smooth height field, our process begins by computing a line field in each slice. This involves calculating the Hessian matrix $\mathbf{H}$, which is defined as
\begin{equation}
\mathbf{H} =
\begin{bmatrix}
\frac{\partial^2 f}{\partial x^2} & \frac{\partial^2 f}{\partial x \partial y} \\
\frac{\partial^2 f}{\partial y \partial x} & \frac{\partial^2 f}{\partial y^2}
\end{bmatrix},
\label{eq:Hess}
\end{equation}
where $f$ represents the height field function, and the elements of the Hessian matrix represent the function's second-order partial derivatives with respect to $x$ and $y$, measuring the rate of change of the gradient of the field, providing the basis for this calculation.

From these matrices, we identify and extract the eigenvector characterized by

\begin{equation}
\mathbf{H} \mathbf{v} = \lambda \mathbf{v},
\label{eq:charact}
\end{equation}

corresponding to the smallest eigenvalue denoted by $\lambda_{\text{min}}$, with $\mathbf{v}_{\text{min}}$ representing the corresponding eigenvector. This eigenvector indicates the direction of minimal curvature within the slice, effectively highlighting the path of least geometric variation. Tracing this direction from any given point on a ridge line guarantees adherence to the ridge, thereby maintaining a consistent path.
The eigenvalues and eigenvectors of the Hessian matrix $\mathbf{H}$ are determined by Jacobi iteration~\cite[Chapter 11.1]{press1992numerical}, ensuring a precise mathematical framework for navigating the ridge line height field.

A ridge aligns with the mid-polyline, delineating the trajectory of the mid-surface. Thus, navigating this ridge through a process informed by the calculated vectors ensures the mid-polyline's precise alignment with the segmented structure's inherent geometry. This principle is crucial, laying the groundwork for subsequent steps aimed at deriving the mid-surface from an aggregation of mid-polylines.

Upon establishing a vector field of minimal curvature in each slice, our next step involves the meticulous tracing of the mid-polyline. It is vital to acknowledge that a single slice might feature multiple \emph{hill ranges} rather than a single one. This complexity necessitates the delineation of connected components within the segmentation, as demonstrated in~\cref{fig:dilation}. For each distinct component, the initiation point for mid-polyline tracing is identified by pinpointing the peak of the smoothed height field—the pixel or voxel exhibiting the maximal value.

With a starting point selected and set as the current point $\mathbf{r}_i$, we perform a computationally cheap Euler integration step to determine the next point $\mathbf{r}_{i+1} = (x_{i+1}, y_{i+1})$ along our mid-polyline calculated as
\begin{equation}
 \mathbf{r}_{i+1} = \mathbf{r}_i + h \cdot \mathbf{v}_i,
 \label{eq:Euler}
\end{equation}
where $h = \sqrt{2} \times \text{spacing}$ is the integration step, \textit{spacing} denotes the distance between slices, effectively defining the separation between mid-polylines, and $\mathbf{v}_i = (x_{v_i}, y_{v_i})$ is the direction of the mid-polyline at the point $\mathbf{r}_i$, corresponding to the eigenvector with the minimal eigenvalue (see~\cref{eq:charact}).
We can avoid more accurate but computationally expensive integration methods since at each step of the integration, the position of the computed point is fine-tuned using the golden section search method~\protect\cite[Chapter 10.1]{press1992numerical}. If the pixel that we are on is not the pixel with the maximum value along the maximal principal direction (perpendicular to the current direction of the line field), we move our point accordingly, as shown in~\cref{fig:mask_to_line_field_2} (F).

Tracing of the mid-polyline continues until it either completes a loop or exits the connected component's boundary. 
When a point exits the connected component and does not make a loop, tracing also proceeds in the opposite direction from the starting point to trace the entire polyline.

A crucial aspect of the tracing process is the treatment of the vector field as a line field, given that our mid-polyline lacks inherent direction. Consequently, vectors are devoid of inherent directionality and are flipped as necessary to ensure continuity. Specifically, we adjust vectors if their orientation deviates by more than $90^{\circ}$ from the preceding vector~\cite{benger2006strategies}.

In cases where the boundary is very thin, the next point in the tracing process might fall outside the boundary, prematurely ending the process. 
To address this, morphological dilation is applied to the connected components, increasing their width by one pixel. 
This process enables a more accurate determination of whether an exit from the boundary is genuine or a result of its thinness, as demonstrated in~\cref{fig:dilation}. 
It is important to note that the dilated slice is used solely to check for the line termination, thus avoiding errors when the boundary is thin.
It is not used to calculate the field and, thereby, does not change the morphological characteristics of the segmentation.  

In scenarios characterized by intricate topologies or suboptimal segmentations—such as branching structures uncommon in biological data but frequently encountered in low-resolution datasets—an additional verification phase is implemented upon the completion of tracing. This phase assesses whether the tracing comprehensively covered the connected component or if specific sections, like branches, were inadvertently missed. Detected untraced segments are then regarded as independent connected components, prompting a re-initiation of the mid-polyline extraction process for each, employing the same methodology as initially outlined.
 
Upon thoroughly connecting all points within a single connected component on the slice, the procedure advances to the next component. This progression continues slice by slice, ensuring all components within a slice are addressed before moving to the next. Following the extraction of mid-polylines from every slice, the workflow progresses to the subsequent phase of our method, namely, the triangulation of these extracted mid-polylines (\cref{sec:meshing}).
The overall process from segmentation to the extraction of the mid-polyline is illustrated in~\cref{fig:mask_to_line_field_2}.

\subsection{Polyline Zipper Algorithm}
\label{sec:meshing} 
We developed a triangulation method named the Polyline Zipper Algorithm to transform the extracted mid-polylines into a mid-surface mesh. 
This algorithm efficiently utilizes the extracted polylines stacked on parallel planes at equidistant levels, where each horizontal level corresponds to a single slice of the volumetric data. At each horizontal level, the mid-polyline comprises either a single continuous line or multiple line segments. The vertices and edges from each pair of adjacent lines (slices) metaphorically act as the \emph{teeth} in the mesh generation process, akin to a zipper.

In our zipper-like approach, vertices from adjacent slices are systematically connected, traversing across the two neighboring polylines and triangulating edges from both sides, resulting in the formation of triangle strips.
This method seamlessly connects all horizontal polylines via vertical edges, ensuring comprehensive connectivity across the resulting triangular surface mesh.

The edges contributing to zipper connectivity fall into two categories.
Two edges, $e_i \in S_i$ and $e_j \in S_{i+1}$, form a valid pair if the Euclidean distance between their centers, $d(e_i, e_j)$, is shorter than the distance between $e_i$ and any other edge $e_k \in S_{i+1}$, and shorter than the distance between $e_j$ and any other edge $e_m \in S_i$. Mathematically, a valid pair is defined by satisfying the following two conditions:
\begin{equation}
d(e_i, e_j) < d(e_i, e_k), \quad \forall e_k \in S_{i+1},
\label{eq:cond01}
\end{equation}
\begin{equation}
d(e_i, e_j) < d(e_m, e_j), \quad \forall e_m \in S_i.
\label{eq:cond02}
\end{equation}

If $e_i$ is found to be closest to $e_j \in S_{i+1}$, but $e_j$ already has another edge $e_m \in S_{i}$ that is closer than $e_i$, then $e_i$ is considered a non-pair edge. In other words, if the condition of \cref{eq:cond01} is satisfied but not that of \cref{eq:cond02}, then  $e_i \in S_i$ is said to be a non-pair edge. Similarly, if  \cref{eq:cond02} is satisfied but not  \cref{eq:cond01}, then  $e_j \in S_{i+1}$ is a non-pair edge. 

\begin{figure}[tb]
    \centering  
   \includegraphics[width=0.8\linewidth]{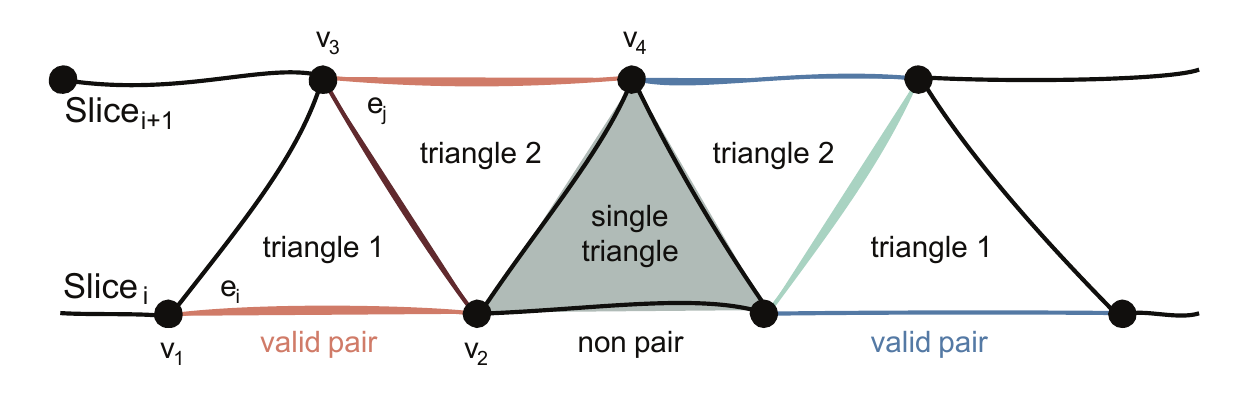}
    \caption{Zipper approach for connecting edges from two neighboring slices. Valid pairs, exemplified by edges $e_i \in Slice_i$ and $e_n \in Slice_{i+1}$, form two triangles (Left). The green-colored triangle represents a single triangle formed by a non-pair edge (center). On the right, the connection of another valid pair of edges is depicted.  }
    \label{fig:2Triangles}
\end{figure}

The four vertices of  $v_1$,  $v_2$, $v_3$, and  $v_4$ of each valid pair of edges are connected to form two triangles (see \cref{fig:2Triangles}). 
To ensure higher-quality triangles, the diagonal edge is established between vertices $v_1$ and $v_3$, or between $v_2$ and $v_3$, prioritizing vertices with larger interior. For example, in \cref{fig:2Triangles} (left valid pair),  $m\angle v_1v_2v_4$ is greater than $m\angle v_3v_1v_2$ so $v_2$ and $v_3$ are diagonally connected. For the non-pair edge  $e_i \in S_i$, the valid vertex from the nearest edge $e_j \in S_{i+1}$ is connected to form a single triangle. 
Moreover, the method takes into account the distance between slices and the absence of edges adjacent to the zipper-like connectivity, which aids in the identification and preservation of holes within the mesh. 
These unconnected regions are intentional, as they accurately represent voids in the original data, ensuring the mesh's integrity and fidelity to the segmented structure.

Triangle strip generation progresses iteratively between two consecutive polylines across the full slice stack. 
Once done, the segmented dataset yields a complete mid-surface mesh.
The process of the mesh generation from the input stack of mid-polylines to the final surface mesh is presented in ~\cref{fig:meshing_process}. 

Unlike reconstruction algorithms such as \ac{sssr}~\cite{boltcheva2016simple} or Delaunay mesh reconstruction~\cite{gong2015approximate}, our approach directly connects the previously extracted points in their precise locations, thereby mitigating the risk of losing the original input geometry's fidelity.
Moreover, our direct triangulation method distinctively handles potential gaps in the data without the need for specifying extra input parameters, setting it apart from conventional mesh reconstruction techniques.
The output mesh not only meets the targeted accuracy requirements but also demonstrates superior mesh quality. Furthermore, this mesh is ready for subsequent refinement processes, including smoothing, simplification, or improvements in mesh quality, according to the specific demands of the application.

\section{Experimental Results}
\label{sec:results}
In this section, we present the outcomes of applying our approach across various datasets and scenarios, demonstrating its versatility and efficacy in mid-surface extraction and analysis.
Our method's performance is first evaluated through a series of surface generation examples in~\cref{sec:surfaceResults}, where we present its capabilities and compare them against those of traditional approaches.
Next, we analyzed the mesh quality visually and numerically and compared our mesh quality with the existing method in~\cref{sec:result:MeshQuality}.
Finally, we illustrate the practical applications of our results in two specific domains: the modeling of biological structures and the estimation of membrane curvature from microscopic data, presented in~\cref{sec:usecasesResults}.
The proposed algorithm is implemented as ParaView~\cite{ahrens2005paraview} plugin in C++ and tested on Intel(R) Xeon(R) Gold 6230R CPU  $2\times 2.10$ GHz with 156 GB RAM and $64$ bit Windows 10 operating system.

For our surface generation experiments, we used a well-documented segmentation dataset provided by Klein~\etal~\cite{klein2020sars}, showing the structural details of SARS-CoV-2. This dataset is particularly useful for showcasing our results because it includes segmentation of the virus and its host cells, which feature membranes of notable thickness. The data was initially segmented through automated processes and then refined manually, providing us with high-quality, reliable data for our initial experiments.  
We used another dataset from Barad~\etal~\cite{barad2023quantifying}, enabling direct comparison with their methodology. 
This dataset consists of segmentation volumes highlighting the mitochondrial structure and contains two distinct labels: the inner and outer mitochondrial membranes. These segmentations were derived semi-automatically~\cite{martinez_sanchez2014robust} from specimens prepared using cryo-FIB milling and scanned using \ac{Cryo-et}.
This dataset, despite its complexity beyond our initial experiments, offers a tangible real-world application scenario for our algorithm, demonstrating its applicability.
\subsection{Surface Generation Results} 
\label{sec:surfaceResults}
\begin{figure}[tb]
    \centering 
    
    
    \includegraphics[width=0.8\linewidth]{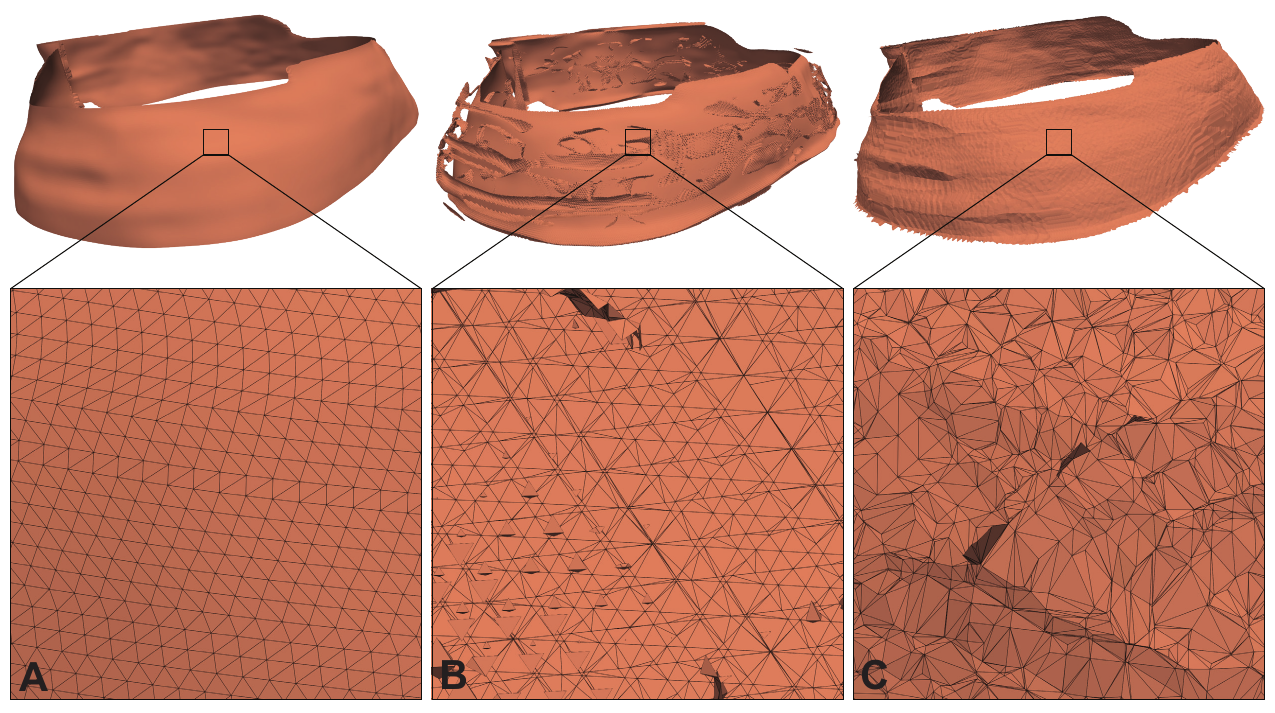}
    \caption{
    Evaluation of different surface types highlighting their limitations for our application. 
    (A) The mid-surface as a result of our method.
    (B) The ridge surface, generated with the VCG ParaView Plugin~\cite{vcg_paraview_plugins}.
    (C) The medial surface as produced by the VoxelCores~\protect\cite{yan2018voxel}.}
    \label{fig:results_other_surfaces}
\end{figure}

The final outcomes of our algorithm are triangular surface meshes. Therefore, to evaluate our algorithm, we used different types of data to observe the visual results of our method and confirm its robustness. First, we examine the applicability of two existing methods for surface extraction. 
As described in~\cref{sec:related-work}, various methods exist for generating surface meshes, including medial and isosurfaces. 
However, they are not directly applicable to our target objectives. 
To highlight these differences and shortcomings, we tested two existing approaches, namely the ParaView plugin~\cite{vcg_paraview_plugins} for ridge surface extraction and the VoxelCores method~\cite{yan2018voxel} for the medial surface extraction. 
The results are shown in~\cref{fig:results_other_surfaces}, where we can see that the mid-surface generated by our method has no specific issue. In contrast, both of the previous methods introduced unnecessary additional surface portions and redundant mesh elements in the output mesh. This underscores the necessity and significance of our algorithm in providing robust results. 

\begin{figure}[htbp]
    \centering
    \includegraphics[width=0.8\linewidth]{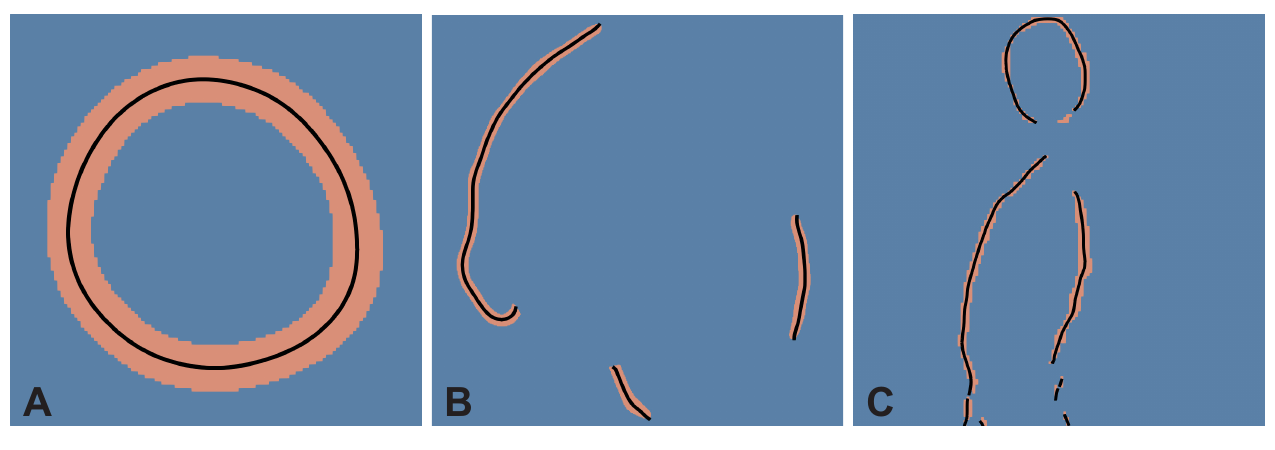}
\caption{Performance of the Mid-Polyline Extraction Algorithm across slices featuring varied topologies. 
(A) A circular segmentation, illustrating the algorithm's efficacy in linking connected loops. 
(B) A slice containing multiple components, resulting in the generation of several mid-polylines. 
(C) A complex segmentation with notably thinner boundaries. 
The segmentations were sourced from datasets provided by Klein~\etal~\protect\cite{klein2020sars} (A, B) and Barad~\etal~\protect\cite{barad2023quantifying} (C).}
    \label{fig:results_midline}
\end{figure}

Next, we evaluated the robustness of our Mid-Polyline Extraction Algorithm. 
\cref{fig:results_midline} shows the results of applying the Mid-Polyline Extraction Algorithm to slices from diverse datasets. These results demonstrate that despite variations in segmentation thickness and complexity, our algorithm consistently performs well.

\begin{figure}[tb]
    \centering
    \includegraphics[width=0.8\linewidth]{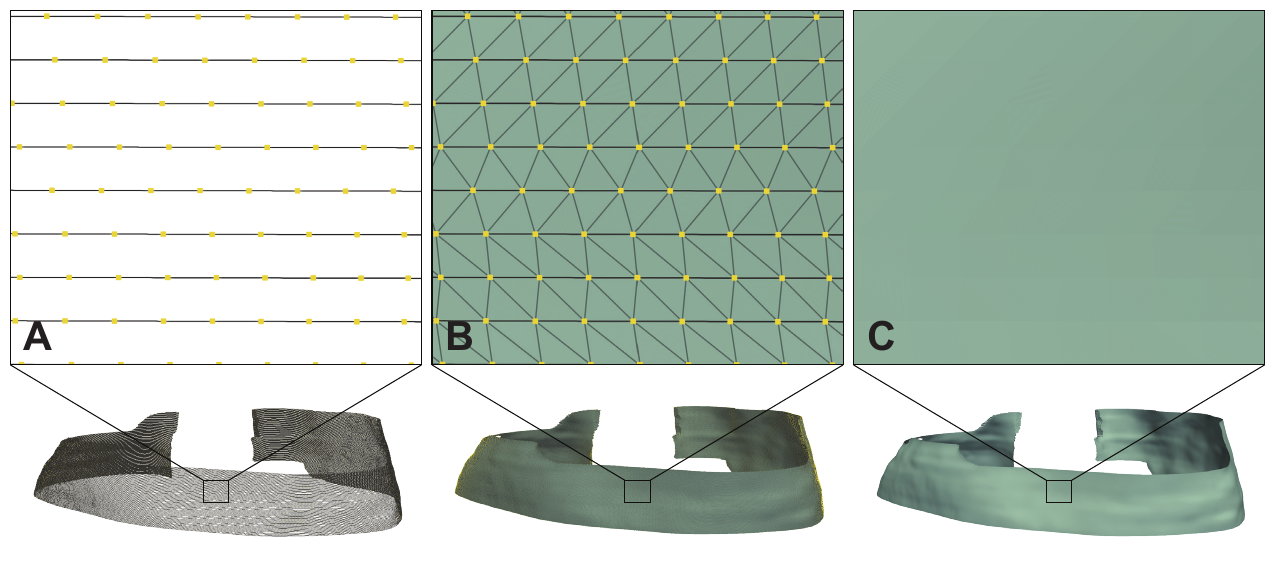}
    \caption{Mesh generation results. 
    (A) Mid-polylines and the points extracted from slices. 
    (B) The triangle mesh, a result of the triangulation process with new edges connecting points between the mid-polylines. 
    (C) The final surface mesh without visible edges to show visual clarity.}  
    \label{fig:meshing_process}
\end{figure}

\begin{figure}[htbp]
    \centering
    \includegraphics[width=\linewidth]{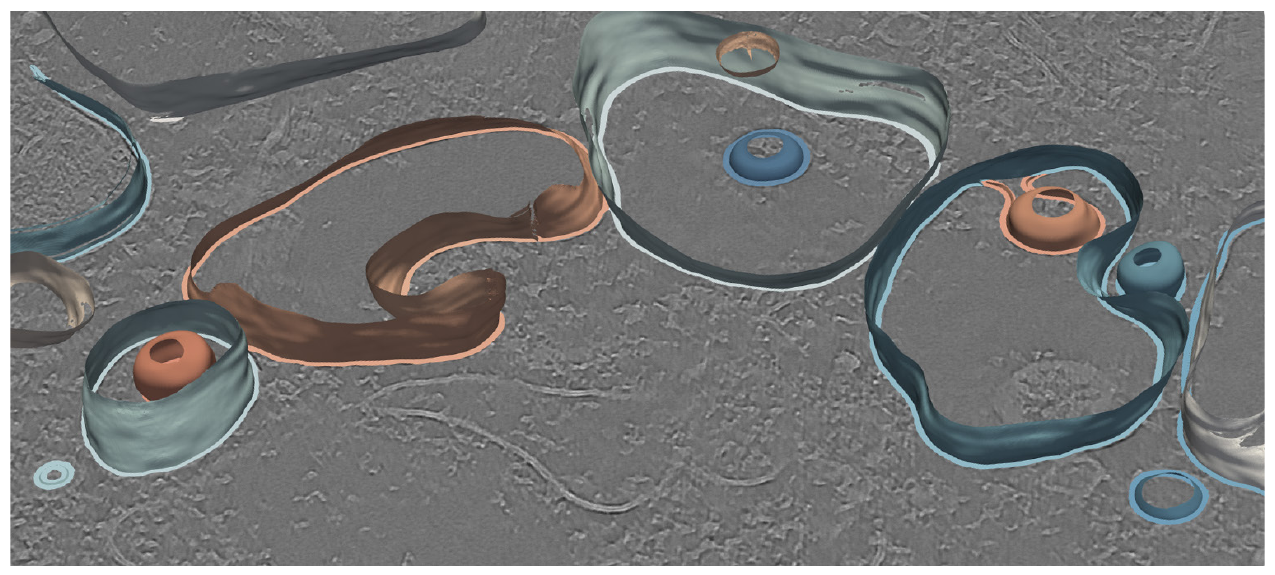}
    \caption{Mid-surfaces generated from the SARS-CoV-2 dataset~\protect\cite{klein2020sars} overlayed on the original tomogram slice and the segmentation mask.}
    \label{fig:results_klein}
\end{figure}

We also evaluated our method on various types of data to see the overall performance from input volumetric data to the final surface mesh. \cref{fig:teaser,fig:meshing_process} show the results of two instances of the SARS-CoV-2 dataset~\cite{klein2020sars}, along with the intermediate results of mid-polylines. The surface meshes generated from the overall dataset~\cite{klein2020sars} are presented in \cref{fig:results_klein}. We can see a slice of the segmentation and the corresponding extracted mid-surface meshes. The results contain $25$ isolated objects. Most of these objects in the segmented input data are without missing information, as exemplified by the three models shown in \cref{fig:teaser,fig:meshing_process,fig:results_other_surfaces}. However, for some objects, the input segmentation has holes in the surfaces, which are preserved in the output. These holes might be true gaps in the surface or missing information due to noise in the data. In either case, our algorithm generates surfaces based on the input data without adding new details, preserving the holes.

\subsection{Assessment of Mesh Quality}
\label{sec:result:MeshQuality}
Mesh quality is essential for various applications. In this section, we evaluate the quality of the generated meshes and compare their visual appearance and quantitative metrics with existing methods. First, we describe the mesh quality metrics used for our analysis, and then we present the results.
\paragraph{Mesh quality metrics used:}
Following standard quality metrics from the meshing domain~\cite{khan2022surface}, we evaluated our mesh using the \textit{triangle quality (Q)}, whose value varies from 0 (poor-quality) to 1 (good quality) and is calculated as:
\begin{equation}
Q= \frac{6}{\sqrt{3}} \frac{A}{(p\times h)},
\label{eq:trianglQty}
\end{equation} 
were, \( A\) is the area of triangle  \( p\) is its half-perimeter, and \( h \) is the length of its longest edge. The angles of the triangles are also crucial for assessing quality. Small and large angles can indicate poor quality. Therefore, we computed the \textit{percentage of angles less than \( 30^\circ \)} and \textit{greater than \( 120^\circ \)}. Similarly, we determined the \textit{average value of the minimal angles} \( \overline{\theta}_{\text{min}} \) in all triangles. Valence refers to the number of adjacent edges to a vertex. While 6 is considered the optimal valence, we noted the \textit{percentage of vertices with valence numbers 5, 6, or 7} $V_{567}$, which is an important metric~\cite{AGHDAII2012v567}. The optimal valence meshes are called regular meshes that are easily remeshed to improve other quality metrics. 

\begin{figure}[htbp]
    \centering
    \includegraphics[width=\linewidth]{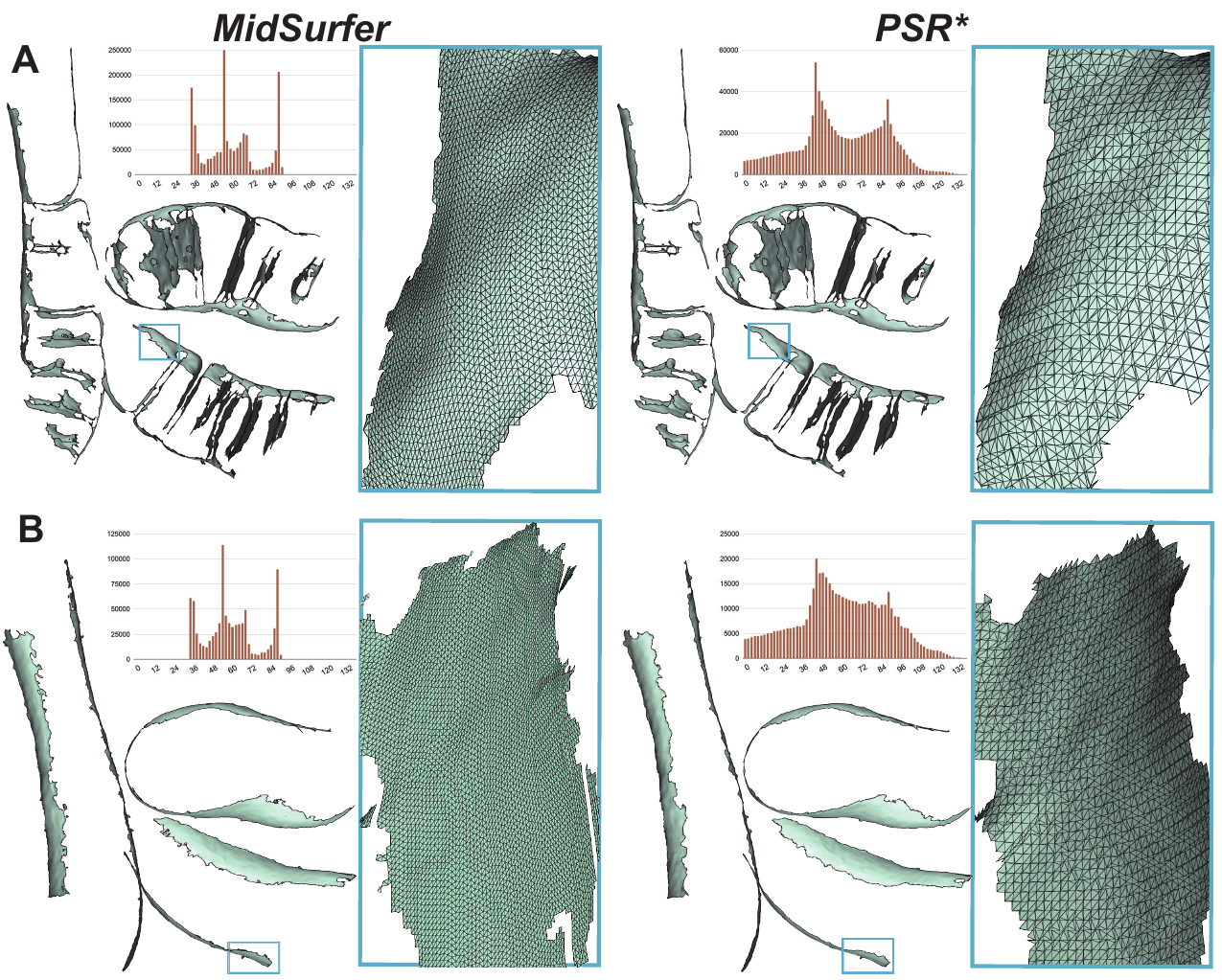}
    \caption{Triangulation results: Left: Ours, Right: \textit{PSR*}, the \ac{PSR}~\protect\cite{kazhdan2013screened} based approach developed by Barad~\etal~\protect\cite{barad2023quantifying}. The top row shows TE7-IMM, and the bottom row depicts TE7-OMM. In the angle histogram, the x-axis indicates angle measurements in degrees, while the y-axis shows the frequency of angles for each value. We observe that our method yields a higher ratio of angles near 60 degrees. 
    The quantitative results are shown in~\cref{tab:mesh_quality}.}
    \label{fig:results_triangulation}
\end{figure}

\paragraph{Mesh quality results:}
\cref{tab:mesh_quality} presents the numerical mesh quality results, which show our meshes are highly regular with a higher ratio of $v567$ vertices. The ratio of poor triangles (lower value of $Q$, triangles with small or large angles) is also very small. 
Let's look into the mesh quality metrics of the existing recently proposed method in the same domain for mid-surface meshes \ie the \ac{PSR}~\cite{kazhdan2013screened} based approach developed by Barad~\etal~\cite{barad2023quantifying}. Our results show a significant improvement. Similarly, the visual results of the meshes are shown in~\cref{fig:results_triangulation}, where, if we look into the zoom view, we can see a significant improvement in mesh regularity and angle quality over the existing method~\cite{barad2023quantifying}. The generation of a high-quality mesh is ensured by generating a highly regular pattern of vertices in the mid-polylines and then triangulating these lines using a quality-aware method by the Polyline Zipper Algorithm. In addition, \cref{tab:mesh_quality} includes a time analysis, indicating that our method demonstrates longer processing times compared to that of Barad~\etal~\cite{barad2023quantifying}, primarily due to the higher number of vertices involved. 
Still, the times are on the same order of magnitude and remain within acceptable limits. Additionally, our method has a high potential for acceleration through parallelism.  The higher number of vertices is due to ensuring high accuracy by considering the same resolution as that of the input data, which defines the inter-slice spacing, thereby affecting the step size in point generation. 
 

\begin{table}[ht]
\setlength{\tabcolsep}{2pt}
\centering
\caption{Quantitative results of mesh quality (PSR*~\protect\cite{barad2023quantifying} vs.\ Ours). $Q$ represents triangle quality calculated via~\cref{eq:trianglQty}. $\boldsymbol{\overline{\theta_{min}}}$ is the average of the minimal angles of all triangles, and $V_{567}$ represents the percentage of regular vertices (having valences 5, 6, or 7). Additionally, the percentages of small and large angles are also shown. The time (in Ours result) is the time taken for mid-polylines generation + the time taken for meshing. \textit{PSR*} time is the total time taken. }
\label{tab:mesh_quality}
\resizebox{0.8\linewidth}{!}{ 
\begin{tabular}{c|c|c|c|c|c|c|c|c|c}
Method&Model& \#Vertices&$Q_{min}$&$\boldsymbol{Q_{avg}}$&$\boldsymbol{\overline{\theta_{min}}}$&
$\boldsymbol{\theta<30^\circ}$&$\boldsymbol{\theta>120^\circ}$ &$\boldsymbol{V_{567}}$&Time (sec.)\\
\hline
\hline
 PSR*&\cref{fig:results_triangulation} (A)& 165808 & 0.00 & 0.60 & 31.69 & 14.00 $\% $ & 0.77 $\%$ & 67 $\%$ & 172\\
Ours&\cref{fig:results_triangulation} (A) & 289088 & 0.03 & 0.78 & 41.26 & 0.02 $\% $  & 0.01 $\%$ & 89 $\%$ & 332 + 130\\
\hline
 PSR*&\cref{fig:results_triangulation} (B) & 87423 & 0.00 & 0.60 & 31.0 & 14.88 $\%$ & 0.91 $\%$ & 70 $\%$ & 92\\
Ours&\cref{fig:results_triangulation} (B) & 150234 & 0.00 & 0.79 & 42.57 & 0.05 $\%$ & 0.02 $\%$ & 92 $\%$ & 283 + 40\\
\hline
Ours&\cref{fig:meshing_process}& 69939 & 0.01 & 0.83 & 44.56 & 0.03$\%$ & 0.01 $\%$ & 97 $\%$ & 14 + 12\\
Ours & \cref{fig:modeling_results} (A) & 10409 & 0.23 & 0.82 & 43.93 & 1.09 $\%$ & 0.0 $\%$ & 97 $\%$ & 2 + 1\\
Ours&\cref{fig:modeling_results} (B) & 39796 & 0.04 & 0.81 & 44.0 & 0.02 $\%$ & 0.01 $\%$ & 97 $\%$ & 4 + 4\\
\end{tabular}%
}
\end{table}

\subsection{Use Case Applications}
\label{sec:usecasesResults}
\begin{figure}[htbp]
    \centering
    \includegraphics[width=0.8\linewidth]{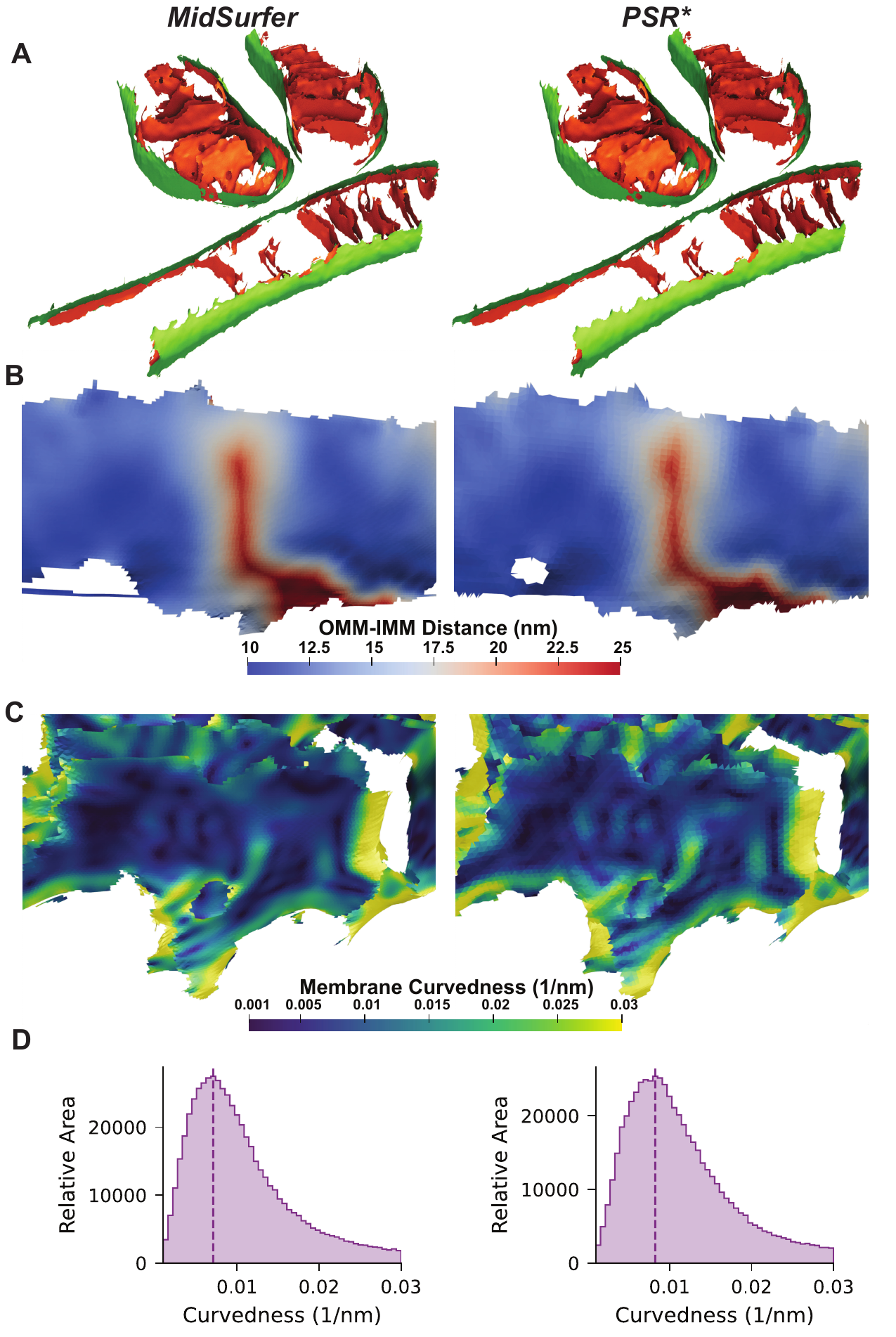}
    \vskip -0.245cm
    \caption{Morphometric comparison of membrane surfaces generated with MidSurfer vs.\ Screened \ac{PSR}. (A) Surface meshes were generated from segmentations depicting mitochondrial inner and outer membranes observed by \ac{Cryo-et}, then quantified with the Surface Morphometrics toolkit~\protect\cite{barad2023quantifying}. Visual comparison shows similar results overall, with improved smoothness of quantification for (B) inter-membrane distance and (C) membrane curvedness, demonstrating the effectiveness of MidSurfer surfaces for downstream quantitative analysis.  (D) Histogram analysis of the surface measurements shows comparable collective quantification of curvedness, but with a smaller high-curvature tail as a result of reducing artifactual high-curvature measurements. \textit{PSR*} is the \ac{PSR}~\protect\cite{kazhdan2013screened} based approach developed by Barad~\etal~\protect\cite{barad2023quantifying}.}
    \label{fig:morphometrics}
\end{figure}
Mid-surface, once extracted, can be utilized in many important applications. 
In this section, we showcase two use cases of the generated mid-surfaces obtained from \ac{Cryo-et} experiments.

\paragraph{Surface morphometrics:}      
 The Surface Morphometrics toolkit~\cite{barad2023quantifying} was recently developed to quantify the ultrastructure of biological membranes, including membrane curvature, orientation, and inter-membrane spacing, using mid-surface models. For this purpose, they utilized the screened PSR method~\cite{kazhdan2013screened} to generate mid-surface meshes, which the toolkit~\cite{barad2023quantifying} then uses for various statistical analyses. Here, we aim to apply our mid-surface meshes for these quantifications and compare them with the method proposed by Barad~\etal~\cite{barad2023quantifying}.

\cref{fig:morphometrics} illustrates the results of these quantifications for mid-surfaces generated using MidSurfer compared to those generated by the Screened PSR-based method~\cite{kazhdan2013screened, barad2023quantifying}. The results are comparable both visually and statistically, with MidSurfer largely reproducing similar results. This underscores that the meshes produced by MidSurfer yield results comparable to those of a recently proposed method. The consistent triangle size and high degree of smoothness slightly reduced high curvature measurements introduced by quantization artifacts in flat membrane segments and generally improved visualization smoothness.

It is important to note that achieving results similar to those presented in \cref{fig:morphometrics} using the PSR-based method~\cite{barad2023quantifying} requires expert users to fine-tune the mid-surface extraction parameters. These results were achieved after the authors of the technique~\cite{barad2023quantifying} invested time in fine-tuning mid-surface extraction parameters to achieve the best possible results. It is highly unlikely that infrequent or new users would achieve similar accuracy. In contrast, MidSurfer employs a parameter-free approach, providing results with a single click. For such users, our approach is likely to yield superior mid-surface extraction compared to manually tuning the Barad~\etal~toolkit~\cite{barad2023quantifying}.
\paragraph{Modeling:}
Mid-surface extraction plays a crucial role in biological modeling and \ac{MD} simulations, where accuracy and realism are paramount. These simulations aim to mimic the behavior of biomolecules and their interactions in a computational environment, providing invaluable insights into biological processes at the molecular level. A key challenge in \ac{MD} simulations is accurately representing lipid bilayer membranes, essential for cell structure and function.

\begin{figure}[t]
    \centering
    \includegraphics[width=0.49\linewidth]{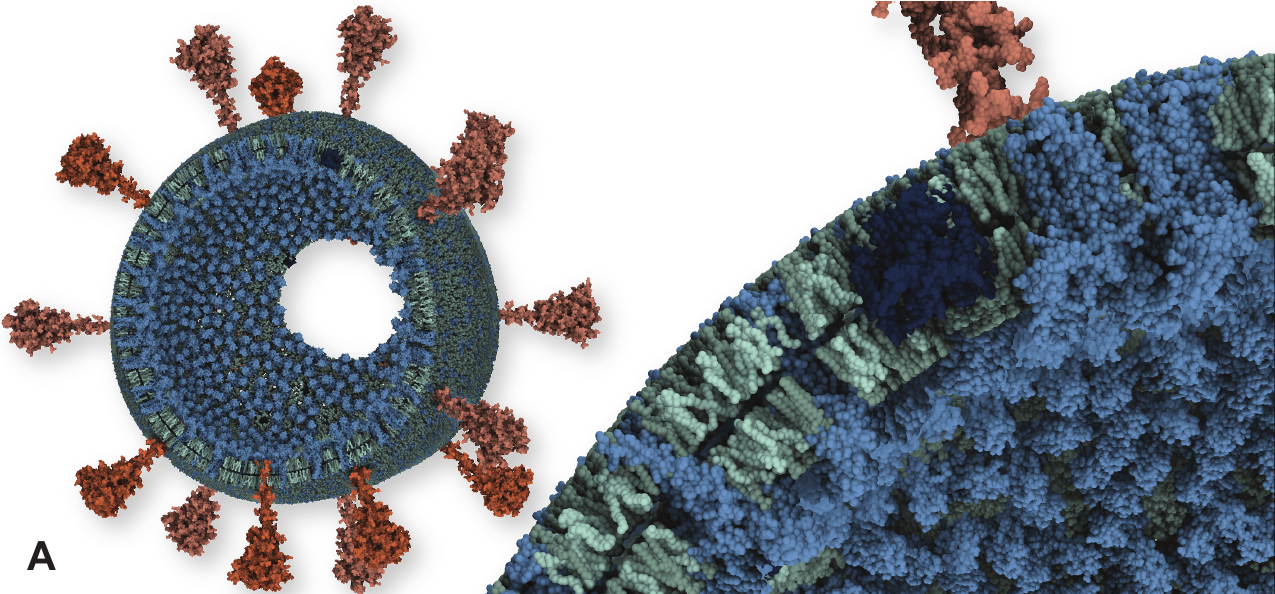}\hfill
    \includegraphics[width=0.49\linewidth]{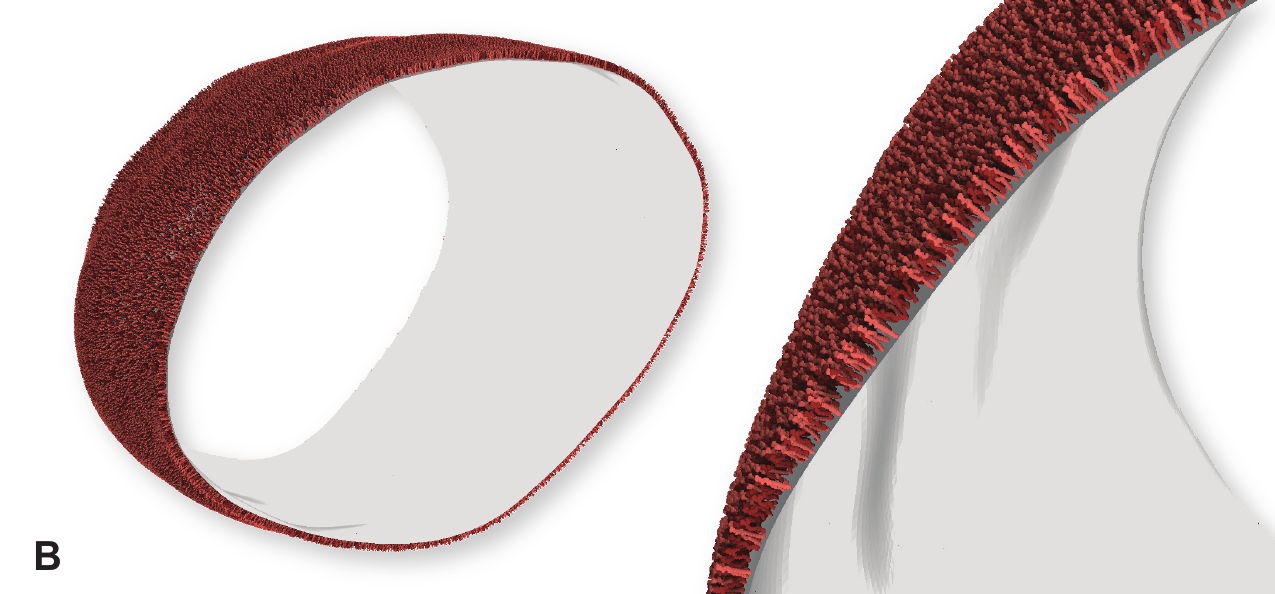}
    \caption{Modeling results with MesoCraft~\cite{nguyen2020modeling}. (A) SARS-CoV-2 virion with the lipids of the lipid bilayer membrane placed on both sides of the extracted mid-surface. (B) Lipids of the lipid bilayer membrane only populated on the outer side of the surface.}
    \label{fig:modeling_results}
\end{figure}

Here, we highlight a use case of mid-surface extraction in biological modeling using MesoCraft, a mesoscale modeling tool~\cite{nguyen2020modeling}. We demonstrate the applied utility of mid-surface extraction within the domain of biological modeling. Focusing on the SARS-CoV-2 virion. As illustrated in~\cref{fig:modeling_results}, we leverage the dataset provided by Klein~\etal~\cite{klein2020sars} for the generation of detailed viral surface depictions. The advantage of mid-surface extraction is evident in the positioning of various components on the virion's surface mesh, notably the lipids constituting the lipid bilayer membrane. It effectively avoids artifacts like membrane shrinking or bloating, ensuring faithful representations of biological membranes' behavior. Such accurate modeling was not achievable with other alternatives, such as using the inner and outer surfaces of the particle membrane.

\section{Discussion}
\label{sec:discussion}
We present a technique for mid-surface extraction from volumetric microscopy data of biological structures, addressing a gap in current methodologies. 
Unlike previous methods, which often require extensive manual tuning of parameters, our findings demonstrate that our approach can accurately extract the mid-surface from segmentation data of various topologies without any user-defined parameterization.

Our research further solidifies the choice of mid-surfaces over more commonly utilized surface representations, a conclusion we show in~\cref{fig:results_other_surfaces}. 
The direct comparison conducted on the same dataset reveals that both isosurfaces and medial surfaces exhibit critical limitations that undermine their utility for our scenario.
Notably, the definitions of isosurfaces and medial surfaces invariably lead to the generation of artifacts that persist despite exhaustive manual adjustments of parameters.
Furthermore, these methods' generation of irregular meshes necessitates subsequent remeshing to attain a usable form for further analysis, presenting a significant procedural inefficiency. 
These findings strongly support the use of the mid-surface approach that not only captures the complex geometries of biological structures but also avoids the common problems associated with more traditional surface representations.

As demonstrated in~\cref{fig:results_midline}, our method for mid-polyline extraction shows consistent robustness across various datasets, with segmentation thickness and topological complexity being the primary variables. 
The key to achieving robustness despite varying segmentation thickness lies in our approach of employing the \acf{SDF} for obtaining the ridge line height field as opposed to other smoothing functions.

Our approach to triangulation enables us to generate well-formed triangles consistently since it progresses through the volume slice by slice, yielding the resulting triangles with the same height. 
By choosing an appropriate size for the integration step in mid-polyline extraction, we can achieve nearly equilateral triangles. 
Furthermore, by imposing a deliberate limitation on the edge lengths within our triangulation process, we adeptly avoid the unintentional closure of gaps present in the original segmentation data. 
These gaps, often critical for the structural and functional understanding of biological systems, are preserved, thereby maintaining the correctness of the biological model.

The MidSurfer method relies on the input of accurately segmented volumetric data. This prerequisite, while ensuring high precision in mid-surface extraction, makes our method dependent on segmentation quality. In cases of suboptimal segmentation, particularly with thin boundaries (\eg missing labeled voxels, unconnected components where there should be continuity or unexpected branches from otherwise smooth boundaries), the efficacy of our approach might be affected. Addressing this issue could entail developing a sophisticated preprocessing stage, but currently, this adjustment is left to the discretion of users based on their specific needs.


Additionally, our method in its current form does not cater to the extraction of mid-surfaces from capped structures, \ie segmentations that feature disk-like shapes on the lower or upper bounds without an inner boundary within a single slice. 
Defining a mid-polyline in these cases becomes challenging due to the lack of information from adjacent slices, leading to uncertainty in determining the correct direction for the mid-surface.
Initially, our focus was on microscopic data, which usually do not include these kinds of topologies. 
This is because the structures we are interested in often extend beyond a single volume's thickness, resulting in slices that contain only the interior regions of interest (\eg mitochondria) without the end caps. 
Consequently, we did not prioritize addressing this specific limitation.
However, our current method automatically identifies and bypasses these structures. 
Future efforts could delve into multi-dimensional slicing techniques, using data from orthogonal planes to enhance the extraction process.

In a few cases, MidSurfer failed to reconstruct sections of the surface at the edges of mitochondrial segmentations (~\cref{fig:morphometrics}), where the segmentations were thinner; however, these edge regions are routinely ignored during quantification due to reduced image and segmentation quality caused by the missing wedge~\cite{salfer2020reliable}. These results broadly show that MidSurfer models are appropriate for detailed quantitative analysis of biological membrane ultrastructure.

One of the key features integrated into our mid-surface extraction algorithm by design is its inherent capability for parallelization. 
While the current implementation operates sequentially, the algorithm's structure allows for the future development of a parallelized version that is expected to accelerate processing speeds significantly. 
This design consideration is vital for efficiently handling the extensive volumetric datasets commonly encountered in biological research and, with it, enhancing the algorithm's practicality for research applications.
We have developed and released a plugin to improve the method's accessibility and usability. 
This plugin, which integrates with ParaView~\cite{ahrens2005paraview}---a widely-used platform for data analysis and visualization---makes it easier for researchers to incorporate our mid-surface extraction technique into their existing workflows. 
By making our algorithm available as a plugin, we eliminate the need for users to navigate through multiple toolboxes, thereby simplifying the research process. 
This effort to enhance accessibility is aimed at encouraging broader adoption of our method across different scientific fields, extending its applicability beyond its initial focus on biological analysis and modeling.

The parameter-free mid-surface reconstructions we generate are able to recapitulate the results obtained by the previous best-in-class screened \ac{PSR} approach with parameters optimized by domain experts. Bypassing the need for parameter optimization will increase the accessibility of midsurface-based analysis approaches for microscopists collecting and segmenting biological membranes since poor parameter selection in previous approaches can result in surfaces that are not suitable for quantification. Beyond accessibility, the increased robustness of the results with variables such as segmentation thickness will improve the degree of automation of pipelines using membrane mid-surface models as a critical step for analysis.


\section{Conclusion}
\label{sec:conclusion}
In this work, we have introduced and formally defined a type of surface, denoted as mid-surface, which is an essential structure for modeling and analysis of 3D microscopy data in structural biology. The concept of mid-surfaces has been notably absent in the visual computing literature until recently, when domain scientists developed a highly effective workflow for mid-surface extraction~\cite{barad2023quantifying}, utilizing screened \acl{PSR}~\cite{kazhdan2013screened}. Unlike their bespoke workflow, our approach is grounded in geometric principles and crucially eliminates the need for parametric tuning. It achieves a quality of output comparable to that of the previous method~\cite{barad2023quantifying} under optimal parameter configurations. The mesh produced by our Polyline Zipper Algorithm is characterized by its nearly equilateral triangles, rendering it highly suitable for a range of analytical and processing tasks.
While it might be possible to derive mid-surfaces through the processing of medial surfaces---by pruning branches---or ridge surfaces---by removing disconnected components---and then applying smoothing and remeshing techniques, such an approach would be significantly more laborious and less efficient compared to our streamlined \emph{one-click} solution.

Admittedly, the success of our method is contingent upon the quality of the initial segmentation. Moving forward, we aim to develop a mid-surface extraction workflow that facilitates quick iterations between segmentation and extraction, allowing for visual assessments and iterative refinements to enhance extraction accuracy. Both the mid-polyline extraction and the zipping algorithms present opportunities for parallelization, suggesting potential for their development into high-performance techniques that deliver immediate results.

As a forthcoming development, MidSurfer---our cutting-edge mid-surface extraction algorithm---will be integrated into surface morphometrics workflows within the realm of structural biology, promising substantial contributions to the field. With the algorithm accessible as a ParaView plugin, we anticipate its adoption across new application areas, solidifying its position as a go-to method for mid-surface extraction in diverse disciplines engaged in volume data analysis.


\section*{Acknowledgment} 
This research was supported by King Abdullah University of Science and Technology (KAUST) (BAS/1/1680-01-01), the KAUST Visualization Core Lab, and the Nadia’s Gift Foundation Innovator Award from the Damon Runyon Cancer Foundation (DRR-65-21 to D.A.G.).
\FloatBarrier

 \bibliographystyle{unsrtnat}

\bibliography{references}
\end{document}